\documentclass[journal]{IEEEtran}

\usepackage{cite}

\ifCLASSINFOpdf
  \usepackage[pdftex]{graphicx}

\else
    \usepackage[dvips]{graphicx}
\fi

\usepackage{amsmath}
\usepackage{algorithmic}
\usepackage{algorithm}
\usepackage{array}
\usepackage{url}
\usepackage{booktabs}
\usepackage{cite}
\usepackage{float}
\usepackage{comment}
\usepackage{amsmath}
\usepackage{mathtools}
\usepackage{url}
\usepackage{bm}
\usepackage{verbatim}
\usepackage{amssymb}
\usepackage{Carrickc,lettrine}
\usepackage{glossaries}
\usepackage{siunitx}
\usepackage{float}
\usepackage{calc}
\usepackage{xcolor}

\usepackage{Carrickc,lettrine}

\makeatletter
\newcommand{\DeclareLatinAbbrev}[2]{%
  \DeclareRobustCommand{#1}{%
    \@ifnextchar{.}{\textit{#2}}{%
      \@ifnextchar{,}{\textit{#2.}}{%
        \@ifnextchar{!}{\textit{#2.}}{%
          \@ifnextchar{?}{\textit{#2.}}{%
            \@ifnextchar{)}{\textit{#2.}}{%
              {\textit{#2.,\ }}}}}}}}%
}
\makeatother
\DeclareLatinAbbrev{\Eg}{E.g}
\DeclareLatinAbbrev{\Ie}{I.e}
\DeclareLatinAbbrev{\etc}{etc}
\DeclareLatinAbbrev{\etal}{et~al}

\newcommand{\ie}{i.e}
\newcommand{\eg}{e.g}

\newcommand{\cph}[1]{\mathrm{e}^{\mathrm{j}#1}}

\newcommand{\lTheta}{\text{Theta II}}

\ifCLASSOPTIONcompsoc
  \usepackage[caption=false,font=normalsize,labelfont=sf,textfont=sf]{subfig}
\else
  \usepackage[caption=false,font=footnotesize]{subfig}
\fi

\ifCLASSOPTIONcaptionsoff
  \usepackage[nomarkers]{endfloat}
 \let\MYoriglatexcaption\caption
 \renewcommand{\caption}[2][\relax]{\MYoriglatexcaption[#2]{#2}}
\fi

\DeclareMathOperator*{\minimize}{minimize}
\DeclareMathOperator*{\subjectto}{subject \ to}
\DeclareMathOperator*{\argmax}{arg\,max}
\DeclareMathOperator*{\argmin}{arg\,min}

\newcommand{\revColor}{blue}
\newcommand{\revision}[1]{#1}

\newacronym{MoM}{MoM}{Method of Moments}
\newacronym{MOR}{MOR}{Model Order Reduction}

\newacronym{RMSE}{RMSE}{Root Mean Square Error}

\begin{document}

\title{On Adaptive Frequency Sampling for Data-driven Model Order Reduction
Applied to Antenna Responses}

\author{Lucas~{\AA}kerstedt,~
        Darwin~Blanco,~
        and~B. L. G. ~Jonsson
	\thanks{Lucas {\AA}kerstedt and B. L. G. Jonsson are with KTH Royal
	Institute of Technology, EECS, 100 44 Stockholm, Sweden (e-mail:
      lucasak@kth.se).}
      \thanks{Darwin Blanco is with Standards \& Technology Ericsson AB, Stockholm,
    Sweden.}}


\maketitle

\begin{abstract}
Frequency domain sweeps of array antennas are well-known to be time-intensive,
and different surrogate models have been used to improve the performance.
Data-driven model order reduction algorithms, such as the Loewner framework and
vector fitting, can be integrated with these adaptive error estimates, in an
iterative algorithm, to reduce the number of full-wave simulations required to
accurately capture the requested frequency behavior of multiport array antennas.
In this work, we propose two novel adaptive methods exploiting a block matrix
function which is a key part of the Loewner framework generating system
approach. The first algorithm leverages an inherent matrix parameter freedom in
the block matrix function to identify frequency points with large errors,
whereas the second utilizes the condition number of the block matrix function.
Both methods effectively provide frequency domain error estimates, which are
essential for improved performance. Numerical experiments on multiport array
antenna S-parameters demonstrate the effectiveness of our proposed algorithms
within the Loewner framework, \revision{where the proposed algorithms reach the
smallest errors for the smallest number of frequency points chosen.}
\end{abstract}



\IEEEpeerreviewmaketitle

\section{Introduction}

\IEEEPARstart{S}{imulating} antenna array responses in the frequency domain is
often time-consuming. For example, in \gls{MoM} one needs to calculate the
impedance matrix and solve a linear system for each frequency point. The total
required time depends on the electrical size, the number of antenna ports, and the
frequency band of the problem. Furthermore, the simulation challenges become
even larger when out-of-band properties are needed \cite{emadeddin_fully_2024}.

One method to speed up the process is to determine fewer frequency points in
\revision{combination with a \gls{MOR} process} \cite{dhaene_adaptive_1995,
de_la_rubia_data-driven_2017}. Methods that address the matrix completion time
include using the Toeplitz structure \cite{fenn_moment_1982} and methods to
solve the linear system fast include iterative methods
\cite{zhao_adaptive_2005}.

\revision{
A key part, on which this work is based on, is the Loewner framework, a method
that has been shown to efficiently approximate solutions to partial differential
equations and nonlinear systems \cite{antoulas_model_2016,
simard_loewner_2021}.} Other algorithms and methods for system approximation
include Cauchy interpolation
\cite{peik_multidimensional_1998,yang_interpolationextrapolation_2007} and
vector fitting \cite{gustavsen_rational_1999,deschrijver_macromodeling_2008}. 
The Loewner framework and vector fitting
are categorized as \emph{non-intrusive} (or
\emph{data-driven}) \gls{MOR} methods \cite{aumann_practical_2023}.
Non-intrusive methods use only the input and output data of a system, and not
any prior physical knowledge of a system \cite{grivet-talocia_passive_2015}. In
this paper, the two data-driven \gls{MOR} methods, the Loewner framework and
vector fitting, are used and compared.

Comparisons between the Loewner
framework and vector fitting have been conducted in
\cite{gosea_algorithms_2021,salarieh_review_2021,aumann_practical_2023, lefteriu_new_2010}. These
papers compare the \gls{MOR} methods on various frequency responses. 
Interestingly, there are no comparisons between the methods for antenna
frequency responses that we have found.
The individual use of vector fitting and Loewner matrix-based system
representation in an antenna context has, however, been carried out before in
\cite{becerra_comparison_2018,gustafsson_unified_2022}, and in
\cite{yuan_method_2022,yuan_removing_2022,jonsson_model_2024}, respectively.
One of the contributions of this paper is a comparison of the Loewner framework
and vector fitting, when specifically applied to array antenna responses.

\revision{
Using a system approximation method in an iterative sense to adaptively choose
frequency samples for the construction of a system model is known as
\emph{adaptive frequency sampling}
\cite{mutonkole_adaptive_2017,mutonkole_multivariate_2017, zhu_adaptive_2022}.
Prior works on adaptive frequency sampling algorithms for non-intrusive
\gls{MOR} include the greedy algorithm by Pradovera
\cite{pradovera_toward_2023}, the scheme by Vuillemin and
Poussot-Vassal \cite{vuillemin_constructive_2021}, and the greedy algorithm by
Cherifi \etal \cite{cherifi_greedy_2022}. These algorithms aim to select as few
frequency samples as possible while still obtaining an accurate system
approximation.

In this paper, we introduce two novel adaptive frequency sampling algorithms
based on the \emph{generating system}-approach from \cite{lefteriu_new_2010}.
These adaptive methods are tested on antenna array responses and compared with
the adaptive methods from \cite{pradovera_toward_2023} and
\cite{vuillemin_constructive_2021}, as well as equidistant distributions using
the Loewner framework and vector fitting. From the comparison, the two novel
frequency sampling algorithms yield the best performance in terms of accuracy as
a function of the number of frequency samples used, on the examined systems.
Additionally, an accurate, novel, error estimator of the system approximation
generated using the Loewner framework is presented. System representations are
associated with an error.  Accurate estimation of the introduced error is
essential in an adaptive frequency sampling scheme. In summary, an improved
framework is presented where frequency points are chosen in an iterative process
with \gls{MoM} to yield an error-controlled system representation of the antenna
response. A different approach to a similar problem is presented in
\cite{de_la_rubia_physics-based_2022}.}

This paper is organized as follows. The theory of scalar vector fitting and the
Loewner framework is shown in Section~\ref{sec:theory}.
Section~\ref{sec:adaptive} introduces the theory of the adaptive sampling
algorithms tested in this paper. In Section~\ref{sec:results}, the adaptive
frequency sampling algorithms are tested on various antenna responses. Lastly,
Section~\ref{sec:conclusions} concludes the paper.

\section{Theory}
\label{sec:theory}

\subsection{Vector Fitting}
Vector fitting is well described in \cite{gustavsen_rational_1999,
grivet-talocia_passive_2015}. We repeat
the main steps in the scalar case for completeness. Let $s\in\mathbb{C}$ be the
complex Laplace variable such that $s = \mathrm{j}\omega$, where $\omega \in
\mathbb{R}$ is the angular frequency.
Vector fitting approximates the scalar function $S(s)$ as a
rational function $h(s)$,
\begin{equation}
    h(s) = \sum_{n=1}^{N}\frac{r_n}{s - a_n} + d +se \mathrm{,}
    \label{eq:vectorFitting}
\end{equation}
where $r_n \in \mathbb{C}$ are the residues, $a_n \in \mathbb{C}$ are the poles,
and $d,e \in \mathbb{C}$ are the asymptotic expansion coefficients. Solving for
the coefficients in \eqref{eq:vectorFitting} is a non-linear problem that is
linearized and solved iteratively by first creating the augmented problem 
\begin{equation}
    \begin{cases}
        \sigma(s) h(s) \approx \sum_{n=1}^{N}\frac{r_n}{s-\tilde{a}_n } + d +se \\
        \sigma(s) \approx \sum_{n=1}^{N} \frac{\tilde{r}_n}{s - \tilde{a}_n} + 1
    \end{cases}
    \mathrm{,}
    \label{eq:vfAugmented}
\end{equation}
where $\tilde{a}_n$ are the starting poles and are set
according to a heuristic scheme. Multiplying $\sigma(s)$ from
\eqref{eq:vfAugmented} with $h(s)$ and rearranging yields
\begin{equation}
    \sum_{n=1}^{N} \frac{r_n}{s - \tilde{a}_n} + d + se -
    h(s)\sum_{n=1}^{N}\frac{\tilde{r}_n}{s - \tilde{a}_n} \approx h(s) \mathrm{.}
    \label{eq:vfMinimize}
\end{equation}
%
%
Given $\{s_i, S(s_i)\}_{i=1}^{N_s}$ and $h(s_i) = S(s_i)$, \eqref{eq:vfMinimize}
yields a linear system that is solved in the least square sense. Solving the
linear system yields the approximative solution to the unknown coefficients,
$r_n, \tilde{r}_n, d, e$.  Given $\{\tilde{r}_n\}$ a new set of poles
$\{\tilde{a}_n\}$ are obtained by solving the eigenvalue problem 
%
%
%
%
\begin{equation}
    \{\tilde{a}_n\} = \mathrm{eig}\left(\tilde{\mathbf{A}} - \tilde{\bm{b}}
    \cdot \tilde{\bm{r}} \right) \mathrm{.}
    \label{eq:vfEigenValue}
\end{equation}
Here, $\tilde{\mathbf{A}} = \mathrm{diag}(\tilde{a}_1,\dots,\tilde{a}_N)$,
$\tilde{\bm{b}} = [1,\dots,1]^{\mathrm{T}}$, and $\tilde{\bm{r}} =
[\tilde{r}_1,\dots,\tilde{r}_N]^{\mathrm{T}}$. The procedure
\eqref{eq:vfMinimize} - \eqref{eq:vfEigenValue} is repeated for a number of
iterations. The final set of poles is inserted into \eqref{eq:vectorFitting},
which leaves a linear problem, solved in the least square sense to
obtain the final set of residues $\{r_n\}$ and the asymptotic coefficients $d$
and $e$. For a generalization to the matrix case, see \cite{grivet-talocia_passive_2015}.


\subsection{The Block Loewner Framework}
With the block Loewner framework \cite{antoulas_chapter_2017,lefteriu_new_2010}, the transfer function matrix
at each sample is used as data.
Given the discrete data set $P = \{s_i, \mathbf{S}(s_i)  \}_{i=1}^{N_s}$,
where $\mathbf{S}(s_i) \in \mathbb{C}^{p \times
m}$, the set is partitioned into two disjoint sets
\begin{equation}
    \begin{cases}
    P_c = \{(\lambda_i, \mathbf{w}_i): i = 1,\dots,k\}\\
    P_r = \{(\mu_j , \mathbf{v}_j): j = 1,\dots,q\} 
    \end{cases}
    \mathrm{,}
    \label{eq:LoewnerPartitioning}
\end{equation}
where
\begin{equation}
    \begin{rcases*}
    \lambda_i = s_i, \quad \quad  \mathbf{w}_i = \mathbf{S}(\lambda_i), \quad \quad i
	= 1,\dots,k \\ \mu_j = s_{k+j}, \quad \mathbf{v}_{j} = \mathbf{S}(\mu_j),
	\quad j = 1,\dots,q \\
    \end{rcases*}
    , \quad k+q = N_s \mathrm{.}
    \label{eq:LoewnerArrayNotationExpl}
\end{equation}
In this paper, we use \emph{alternate splitting} \cite{karachalios_6_2021} for
partitioning the set $P$ into $P_c$ and $P_r$. With alternate splitting, the
frequencies $s_i$ are sorted according to their angular frequency $\omega_i$.
Then, every other frequency is said to belong to $P_c$. The remaining
frequencies are said to belong to $P_r$. 


From the two data sets $P_c$ and $P_r$, the left and right data (or
more precisely, row and column data) are constructed as follows: 
%
%
\begin{equation}
    \begin{rcases*}
	\mathbf{\Lambda} = \mathrm{diag}(\lambda_1,
	\dots, \lambda_k) \otimes \mathbb{I}_m \in \mathbb{C}^{(mk)\times
	(mk)}\\ 
	\mathbf{R} = [\mathbb{I}_m, \dots, \mathbb{I}_m] \in
	\mathbb{C}^{m \times (mk)}  \\ \mathbf{W} = [\mathbf{w}_1, \dots,
	\mathbf{w}_k] \in \mathbb{C}^{p \times (mk)}
    \end{rcases*}
    \mathrm{,}
    \label{eq:LoewnerRightData}
\end{equation}
where $\mathbb{I}_{m}$ is the $m\times m $ identity matrix, and $\otimes$
indicates the Kronecker product.

From the data set $P_r$, the left data (row data) is constructed
\begin{equation}
    \begin{rcases*}
	\mathbf{M} = \mathrm{diag}(\mu_1, \dots,
	\mu_q ) \otimes \mathbb{I}_p \in \mathbb{C}^{(pq)\times (pq)}\\
	\mathbf{L}= [\mathbb{I}_p, \dots,
      \mathbb{I}_p] ^{\mathrm{T}} \in \mathbb{C}^{(pq) \times p} \\
	\mathbf{V}= [\mathbf{v}_1^{\mathrm{T}} , \dots, \mathbf{v}_q^{\mathrm{T}} ] ^{\mathrm{T}} \in
	\mathbb{C}^{(pq) \times m}
    \end{rcases*}
    \mathrm{.}
    \label{eq:LoewnerLeftData}
\end{equation}
For other approaches to constructing the left and right data, \eg,
\emph{tangential} interpolation, see
\cite{antoulas_chapter_2017,lefteriu_new_2010,mayo_framework_2007}.

From the left and right data, the \emph{block Loewner matrix} is constructed
\begin{equation}
    \mathbb{L} = 
    \begin{bmatrix}
	\frac{\mathbf{v}_1 - \mathbf{w}_1}{\mu_1 - \lambda_1}  & \cdots &
	\frac{\mathbf{v}_1 - \mathbf{w}_k}{\mu_1 - \lambda_k} \\ \vdots & \ddots
									& \vdots
	\\ \frac{\mathbf{v}_q - \mathbf{w}_1}{\mu_q - \lambda_1} & \cdots &
	\frac{\mathbf{v}_q - \mathbf{w}_k}{\mu_q - \lambda_k}
    \end{bmatrix}
    \in \mathbb{C}^{(pq) \times (mk)} \mathrm{.}
    \label{eq:LoewnerMatrix}
\end{equation}

Within the Loewner framework, there exist at least three approaches for
constructing \revision{a system representation} $\mathbf{H}(s)$, such that
$\mathbf{H}(s) \approx \mathbf{S}(s)$. For the first approach \cite[Sec
4]{mayo_framework_2007}, the surrogate model $\mathbf{H}(s)$ is constructed
using the state space representation
\begin{equation}
	\mathbf{H}(s) = \mathbf{C}\left(s\mathbb{I} - \mathbf{A}\right)^{-1}
	\mathbf{B} + \mathbf{D} \text{,}
    \label{eq:LoewnerInterpolant1}
\end{equation}
where the state space matrices $\mathbf{A},\mathbf{B},\mathbf{C},\mathbf{D}$ are
given by
\begin{equation}
    \left( \begin{array}{c|c}
       \mathbf{A} & \mathbf{B} \\
       \midrule
       \mathbf{C} & \mathbf{D} \\
    \end{array}\right)
     = 
     \left( \begin{array}{c|c}
       \boldsymbol{\Lambda} + \mathbb{L}^{\#} (\mathbf{V} - \mathbf{L} \mathbf{D}) & \mathbb{L}^{\#}(\mathbf{V} - \mathbf{L}\mathbf{D}) \\
       \midrule
       -(\mathbf{W} - \mathbf{D}\mathbf{R}) & \mathbf{D} \\
    \end{array}\right)
    \text{,}
    \label{eq:interpolant2StateSpace}
\end{equation}
where $\mathbf{D}$ is arbitrary, and $\mathbb{L}^{\#}$ is the right inverse of
$\mathbb{L}$ (and can be calculated by, \eg, the Penrose-Moore inverse).

The second method of constructing the surrogate model is called the \emph{generating
system} approach \cite{lefteriu_new_2010,antoulas_chapter_2017}. Here,
the block matrix function $\mathbf{\Theta}(s)$, or its inverse
$\mathbf{\bar{\Theta}}(s)$, is
constructed to generate the surrogate model:
\begin{equation}
    \begin{aligned}
    \mathbf{H}(s) = \left[ \mathbf{\Theta}_{11}(s) \mathbf{G}_1(s) \right. &
    \left. - \mathbf{\Theta}_{12}(s) \mathbf{G}_2(s)  \right] \\ & \left[
    -\mathbf{\Theta}_{21}(s) \mathbf{G}_1(s) + \mathbf{\Theta}_{22}(s)
\mathbf{G}_2(s)  \right]^{-1} \text{,}
    \end{aligned}
    \label{eq:LoewnerGenerating}
\end{equation}
or
\begin{equation}
    \begin{aligned}
    \mathbf{H}(s) = \left[  \mathbf{G}_1(s) \mathbf{\bar{\Theta}}_{11}(s)
    \right. & \left. + \mathbf{G}_2(s) \mathbf{\bar{\Theta}}_{21}(s)  \right] \\
	    & \left[ \mathbf{G}_1(s) \mathbf{\bar{\Theta}}_{12}(s)  +
	    \mathbf{G}_2(s)\mathbf{\bar{\Theta}}_{22}(s)   \right]^{-1} \text{,}
    \end{aligned}
    \label{eq:LoewnerGeneratingBar}
\end{equation}
where $\mathbf{G}_1(s)$ and $\mathbf{G}_2(s)$ are arbitrary polynomial matrices
\cite{antoulas_chapter_2017}. If $k = q$, and $\mathbb{L}$ is invertible, the
block matrices $\mathbf{\Theta}(s)\in\mathbb{C}^{(p+m)\times(p+m)}$ and $\mathbf{\bar{\Theta}}(s) \in
\mathbb{C}^{(p+m)\times (p+m)}$ are defined as
\cite{lefteriu_new_2010,antoulas_chapter_2017}
\begin{equation}
    \begin{aligned}
        \mathbf{\Theta}(s) = 
        \begin{bmatrix}
        \mathbb{I}_{p} & 0\\
        0 & \mathbb{I}_{m}
        \end{bmatrix}
        + \begin{bmatrix}
            \mathbf{W} \\
            -\mathbf{R}
        \end{bmatrix}
      (s \mathbb{L} - &\mathbb{L} \mathbf{\Lambda})^{-1} \begin{bmatrix}
            \mathbf{L} & \mathbf{V}
        \end{bmatrix}
        \\
        = 
        & \begin{bmatrix}
            \mathbf{\Theta}_{11}(s) && \mathbf{\Theta}_{12}(s) \\
            \mathbf{\Theta}_{21}(s) && \mathbf{\Theta}_{22}(s) \\
        \end{bmatrix}
        \text{,} 
    \end{aligned}
    \label{eq:loewnerTheta1}
\end{equation}
and
\begin{equation}
    \begin{aligned}
        \mathbf{\bar{\Theta}}(s) = 
        \begin{bmatrix}
        \mathbb{I}_{p} & 0\\
        0 & \mathbb{I}_{m}
        \end{bmatrix}
        + \begin{bmatrix}
            -\mathbf{W} \\
            \mathbf{R}
        \end{bmatrix}
	(s \mathbb{L} - &\mathbf{M}\mathbb{L})^{-1} \begin{bmatrix}
            \mathbf{L} & \mathbf{V}
        \end{bmatrix}
        \\
         = 
        & \begin{bmatrix}
            \mathbf{\bar{\Theta}}_{11}(s) && \mathbf{\bar{\Theta}}_{12}(s) \\
            \mathbf{\bar{\Theta}}_{21}(s) && \mathbf{\bar{\Theta}}_{22}(s) \\
        \end{bmatrix}
        \text{.} 
    \end{aligned}
    \label{eq:loewnerTheta2}
\end{equation}
%
%

The third approach is the barycentric interpolation representation
\cite{antoulas_chapter_2017}
\begin{equation}
    \begin{aligned}
	    \mathbf{H}(s) = \frac{\sum_{i = 1}^{k} \frac{b_i
	    \mathbf{w}_i}{s - \lambda_i}}{ \sum_{i=1}^{k} \frac{b_i}{s -
	\lambda_i}}
	    \text{,}
    \end{aligned}
    \label{eq:loewnerBarycentric}
\end{equation}
where $b_i$ are the barycentric coefficients. The barycentric coefficients $b_i$
are obtained by solving the optimization problem
\cite{pradovera_toward_2023}
\begin{equation}
    \begin{aligned}
	    \minimize \sum_{j=1}^{q}  & \Bigg|\Bigg|\sum_{i = 1}^{k} b_i
	    \frac{\mathbf{v}_j - \mathbf{w}_i}{\mu_j -
	    \lambda_i}\Bigg|\Bigg|^2_{F} \\
				      &  \subjectto \sum_{i=1}^{k} |b_i|^2 = 1
				      \text{,}
    \end{aligned}
    \label{eq:loewnerBarycentricCoeff}
\end{equation}
where $||.||_{F}$ is the Frobenius norm.

\section{Frequency Sampling Algorithms}
\label{sec:adaptive}
Frequency domain simulations of antennas yield the frequency response calculated
at a given discrete set of frequencies. The duration for calculating the frequency
response for one frequency point is often quite high for arrays. It is,
therefore, desirable to keep the number of frequency points low while still
having enough frequency points for the system approximation to accurately
resemble the frequency response. Choosing these frequency points in the given
frequency band is in this work denoted as \emph{frequency sampling},
\cite{dhaene_adaptive_1995,pradovera_toward_2023}. 

Frequency points may either be sampled in an adaptive or predetermined manner.
Predetermined frequency sampling means that a set of frequency points is chosen
prior to any frequency domain calculations. 

In adaptive frequency sampling, an initial set of frequency samples is
iteratively enriched based on an error estimation over a desired frequency band.
First, the error estimation is carried out using the available sample points.
Then, the frequency response from the surrogate model is examined to determine
the largest error and the corresponding frequency point is then added to the
available samples. The procedure then continues iteratively until 
it has reached an estimated error smaller than a given error tolerance or a
maximum number of iterations. Some error estimation methods utilize properties
of the system representation method, whereas some methods only utilize the data
of the constructed model. A general description of the adaptive frequency
sampling algorithms presented here is displayed in Algorithm~1.



The number of frequency points in the initial set varies. In
\cite{yang_accurate_2016}, a sparse uniformly sampled set is used, whereas in
\cite{pradovera_toward_2023}, only one to two frequency points are used in the
initial set. In this work, we use only two frequency points in the initial set:
the lower and upper-frequency limit of the band of interest.

Throughout this paper, we denote the number of samples \emph{used} for system
approximation with $N_s$. 
Once the surrogate model is constructed, it can provide a system approximation
on a fine grid.
This fine grid
consists of $M_s$ equidistantly spaced frequency samples in the frequency band
of interest, $[f_{\mathrm{min}}, f_{\mathrm{max}}]$. Here, we denote these fine
sampled points with $s_{\ell}'$, whereas the samples chosen in an
adaptive or predetermined manner are denoted with $s_{i}$. Typically, the chosen
samples form a subset 
in the band of
interest, \ie, $\{s_i\}_{i = 1}^{N_s} \subseteq \{s'_{\ell}\}_{\ell=1}^{M_s},
N_s \leq M_s$, often $N_s \ll M_s$.

 \begin{algorithm}
 \caption{General adaptive frequency sampling algorithm}
 \begin{algorithmic}[1]
 \renewcommand{\algorithmicrequire}{\textbf{Input:}}
 \renewcommand{\algorithmicensure}{\textbf{Output:}}
 \REQUIRE Initial set of $N$ samples $\{s_i, \mathbf{S}(s_i)\}_{i=1}^{N}$\\ 
          Desired error tolerance: $E_{\mathrm{tol}}$\\
	  Number of maximum iterations: $N_{\mathrm{max}}$ \\
	  Number of points for surrogate model construction:
	  $M_s$ \\
	  Band of interest: $f_{\mathrm{min}}$, $f_{\mathrm{max}}$
 \ENSURE  $\mathbf{H}(s_{\ell}')$
 \STATE Set of samples $\mathcal{S} = \{s_i, \mathbf{S}(s_i)\}_{i=1}^{N}$
 \STATE Points for surrogate model 
 $\{s_{\ell}'\}_{\ell = 1}^{M_s}$ = $2\pi\mathrm{j}$ linspace($f_{\mathrm{min}}$,
	  $f_{\mathrm{max}}$, $M_s$)
 \STATE Initial estimated error: $E_{\mathrm{est}} = \infty$
 \FOR {$n = 0,\dots,N_{\mathrm{max}}-1$}
  \STATE Construct model $\mathbf{H}(s_{\ell}')$ with the $N + n$ samples
  available, for $\ell = 1,\dots,M_s$
  \STATE Estimate error $E_{\mathrm{est}}$ of $\mathbf{H}(s_{\ell}')$
  \IF {$E_{\mathrm{est}}<E_{\mathrm{tol}}$}
  \STATE Break
  \ENDIF
  \STATE Choose next point $s_{N+n+1}$ from the estimated error or by
  analyzing $\mathbf{H}(s_{\ell}')$ (or other system approximation method
  specific properties)
  \STATE Calculate $\mathbf{S}(s_{N+n+1})$ 
  \STATE Add sample: $\mathcal{S} \leftarrow \mathcal{S}
  \cup \{s_{N+n+1}, \mathbf{S}(s_{N+n+1})\}$
  \ENDFOR
 \RETURN $\mathbf{H}(s_{\ell}')$
\end{algorithmic} 
\end{algorithm}

For predetermined frequency distributions, any arbitrary distribution of
frequency points may be used as long as the choice of points is made before any
system approximation is carried out. \revision{In this work, we use an equidistant
distribution. While an equidistant distribution is not practical in an
iterative setting, it serves as a good benchmarking reference.}

Each of the proposed methods below, see Section~\ref{sec:Vuillemin}~-~\ref{sec:ThetaII}, utilize different methods to predict the largest error. To
compare and quantize the performance of these methods, an error between the
constructed surrogate models and the true data is calculated. In this paper, all
system approximation and sampling methods are \emph{evaluated} by the \gls{RMSE},
similarly to \cite{aumann_practical_2023},
\begin{equation}
  \mathrm{RMSE} = \left( \frac{1}{M_s} \sum_{\ell=1}^{M_s}
    \big|\big|\mathbf{S}(s_{\ell}') - \mathbf{H}(s_{\ell}')\big|\big|_F^2
  \right)^{\frac{1}{2}} \text{,}
  \label{eq:RMSE}
\end{equation}
where $\mathbf{S}(s_{\ell}')$ is the testing data, and $\mathbf{H}(s_{\ell}')$
is the constructed surrogate model. Throughout Section~\ref{sec:results}, the
testing data consists of scattering parameters calculated for a discrete set of
frequencies, $s_1',\dots,s_{M_s}'$, obtained using commercial full-wave
simulation software or an in-house \gls{MoM} solver. Therefore, the surrogate
models are constructed for the same set of discrete frequencies,
$s_1',\dots,s_{M_s}'$, to relieve the use of \eqref{eq:RMSE}. 

In Section~\ref{sec:results} we also display the relative error between the
testing data and the surrogate models, as a function of frequency. Here, we consider
the Frobenius norm of the element-wise relative error, $E_{\mathrm{rel}}$, 
\begin{equation}
  E_{\text{rel}}(s_{\ell}') = || \mathbf{E}(s_{\ell}')||_{F}\text{,} \quad
  \mathrm{E}_{ij}(s_{\ell}') = \frac{ \mathrm{H}_{ij}(s_{\ell}') -
    \mathrm{S}_{ij}(s_{\ell}')
  }{\mathrm{S}_{ij}(s_{\ell}') + \delta } \text{,}
	\label{eq:error}
\end{equation}
where $\delta = 10^{-15}$.

\subsection{Double-sided Sampling}
The transfer function $\mathbf{S}(s)$ of an LTI system is \emph{Positive Real}
(the admittance or impedance representation) or \emph{Bounded Real} if and only
if the LTI system is passive \cite[Th 2.1]{grivet-talocia_passive_2015}. Both a
positive real and a bounded real transfer function $\mathbf{S}(s)$ fulfills
\begin{equation}
  \mathbf{S}(s)^* = \mathbf{S}(s^*) \text{.}
  \label{eq:doubleSampling}
\end{equation}
Given $N_s$ discrete samples on the imaginary axis, $\{\mathrm{j}\omega_i,
\mathbf{S}(\mathrm{j}\omega_i)\}_{i=1}^{N_s}$, of a passive LTI system,
the passivity may be exploited to obtain additional $N_s$ samples.
First, consider the sampled set $\{\mathrm{j}\omega_i,
\mathbf{S}(\mathrm{j}\omega_i)\}_{i=1}^{N_s}$. Using \eqref{eq:doubleSampling},
with the already sampled set, we obtain the set $\{-\mathrm{j}\omega_i,
\mathbf{S}(\mathrm{j}\omega_i)^*\}_{i=1}^{N_s}$. This procedure leaves us with the
double amount of samples and is here denoted as \emph{double-sided sampling}.
In this work, we use double-sided sampling unless stated otherwise.

\subsection{Vuillemin Adaptive Frequency Sampling}
\label{sec:Vuillemin}
The adaptive frequency sampling algorithm described in
\cite{vuillemin_constructive_2021} is a heuristic scheme that samples the next
frequency point according to the strongest dynamic of the current surrogate model.
This method identifies the dynamics by finding the peaks and valleys of the
current surrogate model. In \cite{vuillemin_constructive_2021}, the method is used
together with the Loewner framework.

Considering the general description of adaptive frequency sampling in
Algorithm~1, the method in \cite{vuillemin_constructive_2021} specifies line 10,
\ie, the analysis of the surrogate model (or the properties of the system
approximation method) to decide the next point to sample. From the current
surrogate model, $\mathbf{H}(s_{\ell}')$, a scalar function is defined 
\begin{equation}
  g(\omega_{\ell}) = || \mathbf{H}(\mathrm{j} \omega_{\ell})||_2 \text{,}
	\label{eq:constrFunction}
\end{equation}
where $||.||_2$ is the $\ell^2$-norm. The discrete frequency corresponding to
the largest or lowest value of $g(\omega_{\ell})$ is then chosen as the
candidate point. If there are no peaks or valleys, the frequency point
corresponding to the largest or smallest discrete derivative of
$g(\omega_{\ell})$ is selected as the candidate point.

\subsection{Pradovera Adaptive Frequency Sampling}
\label{sec:Pradovera}
Another adaptive sampling algorithm to investigate is the greedy sampling
algorithm described by Pradovera in \cite{pradovera_toward_2023}. In
\cite{pradovera_toward_2023}, an adaptive sampling algorithm is constructed
based on the barycentric interpolant representation (see
\eqref{eq:loewnerBarycentric}) for the Loewner framework. Pradovera shows that
the \emph{relative residual norm} $\rho(s)$ of the surrogate model
$\mathbf{H}(s)$ satisfies \cite[eq.~(10)]{pradovera_toward_2023}
\begin{equation}
	\rho(s) = \gamma \Bigg| \sum_{i = 1}^{k} \frac{b_i}{s-\lambda_i}
	\Bigg|^{-1}
	\text{,}
	\label{eq:pradoveraResidual}
\end{equation}
where $\gamma$ is a frequency-independent constant. Subsequently, the next
frequency point is sampled to minimize the magnitude of the norm of the sum in
\eqref{eq:pradoveraResidual}. A frequency sampling rule is thus obtained:
\begin{equation}
	s_{N_s+1} = \argmin_{s} \Bigg| \sum_{i=1}^{k} \frac{b_i}{s-\lambda_i}
	\Bigg| \text{.}
	\label{eq:pradoveraRule}
\end{equation}

In \cite{pradovera_toward_2023}, double sided sampling is used such that $P_c =
\{s_i,\mathbf{S}(s_i)\}_{i=1}^{N_s}$, $P_r = \{s_i^*,
[\mathbf{S}(s_i)]^*\}_{i=1}^{N_s}$.


\subsection{Loewner Generating System Based Adaptive Sampling I}
\label{sec:ThetaI}
Here, we propose the first of our two novel adaptive sampling methods.
In \cite{lefteriu_new_2010}, an adaptive sampling scheme based on the Loewner
generating system approach is described. The method from
\cite{lefteriu_new_2010} is employed in a context where a large number of
samples are already available, \eg, measured S-parameters. There, the goal is to
add a number of (already available) samples in each iteration to be used with the
Loewner framework to create a reduced order model. 

Recall that in this paper, no samples are available in advance. The choice of
the next frequency point to sample is driven by the already sampled frequency
points. Similarly to \cite{lefteriu_new_2010}, we may, however, use the
generating system approach to create an adaptive sampling scheme. The underlying
philosophy is as follows: 

\begin{figure*}
\centering
\includegraphics[width =\textwidth]{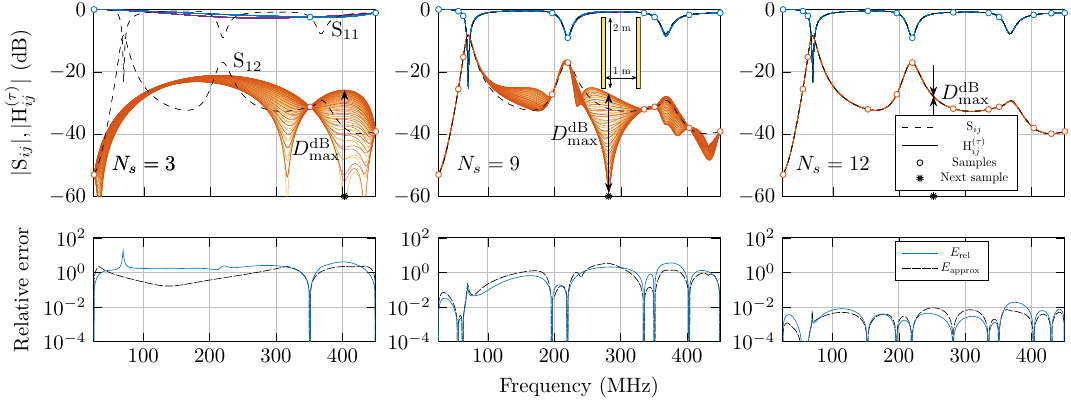}
\caption{Forty surrogate models constructed using \eqref{eq:LoewnerGenerating}
  with the unitary matrices $\mathbf{G}_1^{(\tau)}$ and $\mathbf{G}_2^{(\tau)}$,
  $\tau = 1,\dots,40$, for 3, 9, and 12 available samples. The largest relative
  deviation is highlighted with $D_{\mathrm{max}}^{\mathrm{dB}}$, which decides
  the next frequency point to sample. In the lower plots, the corresponding
  relative error \eqref{eq:error} and the estimated relative error using
  \eqref{eq:errorApprox}, are displayed.}
\label{fig:testFig}
\end{figure*}

$N$ surrogate models are constructed using the generating function
$\mathbf{\Theta}(s)$ (or $\bar{\mathbf{\Theta}}(s)$, see
\eqref{eq:loewnerTheta1} and \eqref{eq:loewnerTheta2}) together with a set of
different arbitrary matrices $\mathbf{G}_1^{(\tau)}$ and
$\mathbf{G}_2^{(\tau)}$, $\tau = 1,2,\dots,N$. A surrogate model constructed
from the matrices $\mathbf{G}_1^{(\tau)}$ and $\mathbf{G}_2^{(\tau)}$ is in this
paper denoted by $\mathbf{H}^{(\tau)}(s_{\ell}')$. The discrete frequency point
for which the surrogate models $\mathbf{H}^{(1,\dots,N)}(s_{\ell}')$
\emph{differ} the most is then selected as the next frequency point to sample.
The difference among the models can be calculated in a number of ways. In this
paper, the element-wise relative difference is calculated, \ie,
\begin{equation}
  s_{N_s +1} = \argmax_{\ell,i,j,\tau,\nu} \Bigg|
  \frac{\mathrm{H}_{ij}^{(\tau)}(s_{\ell}')
  - \mathrm{H}_{ij}^{(\nu)}(s_{\ell}')}{\mathrm{H}^{(\tau)}_{ij}(s_{\ell}')} \Bigg| \text{.}
	\label{eq:relativeDifference}
\end{equation}
It is, however, possible to perform other sorts of difference calculations, \eg,
absolute difference.

The Theta I method relies on the block matrix function $\mathbf{\Theta}(s)$ (or
$\bar{\mathbf{\Theta}}(s)$), which is only defined for when $k=q$ and when
$\mathbb{L}$ is invertible. As a consequence, the method must be fed an even
number of square matrix samples $\{s_i, \mathbf{S}(s_i)\}_{i=1}^{N_s}$ unless we
modify the calculation of the block matrix $\mathbf{\Theta}(s)$. Here, we
propose three ways of managing an odd number of samples. Either use a
pseudo-inverse for the calculation of $(s\mathbb{L} -
\mathbb{L}\mathbf{\Lambda})^{-1}$, or Split the $N_s$ ($N_s$ being odd)
available samples into two groups such that the first group consists of samples
$1,2,\dots,N_s-1$, and the second group consists of samples $2,3,\dots,N_s$.
The calculation in \eqref{eq:relativeDifference} is then performed on the two
groups separately, which yields us two candidate points. Of the two candidate
points, the one corresponding to the largest difference among the corresponding
surrogate models is chosen. Lastly, the principle of double-sided sampling may
be used to always have $2N_s$ samples available. This solution, however, limits
us to passive LTI systems only.

%
%
%

With our proposed method, we may \emph{approximate} the relative error
$E_{\text{rel}}$ by calculating the maximum relative difference among the
surrogate models:
\begin{equation}
  \begin{aligned}
    E_{\text{approx}}(s_{\ell}') = &||\tilde{\mathbf{E}}(s_{\ell}')||_F \text{,}\\
				  & \tilde{\mathrm{E}}_{ij}(s_{\ell}') =
				  \max_{\tau,\nu} \frac{1}{p}\Bigg|
  \frac{\mathrm{H}_{ij}^{(\tau)}(s_{\ell}') -
  \mathrm{H}_{ij}^{(\nu)}(s_{\ell}')}{\mathrm{H}^{(\tau)}_{ij}(s_{\ell}') + \delta} \Bigg| \text{.}
\end{aligned}
	\label{eq:errorApprox}
\end{equation}

In Fig.~\ref{fig:testFig}, we display an example of how the adaptive method is
used to approximate a given antenna response. Here, the Theta I method is
used to approximate the scattering data originating from two \SI{2}{\meter} long
dipole antennas situated \SI{1}{\meter} apart (see inset of
Fig.~\ref{fig:testFig}), operating in the frequency band
\SI{25}{\mega\hertz} to \SI{450}{\mega\hertz}, with \SI{50}{\ohm} port
impedances. For demonstrative purposes, we have chosen our set of matrices
$\mathbf{G}_1^{(\tau)}$, and $\mathbf{G}_2^{(\tau)}$ as unitary $2 \times 2$
matrices on the form
\[
  \begin{aligned}
  & \mathbf{G}_1^{(\tau)} = 
  \begin{bmatrix}
    w & z \\
    -z^* \cph{\theta_{\tau}} & w^*\cph{\theta_{\tau}} \\
  \end{bmatrix}
  \text{,} \quad 
  \mathbf{G}_2^{(\tau)} = 
  \begin{bmatrix}
    x & y \\
    -y^* \cph{\theta_{\tau}} & x^*\cph{\theta_{\tau}} \\
  \end{bmatrix}
  \text{,}
  \\
  & x = \frac{1}{\sqrt{2} }\cph{\varphi_1}, \quad y = \frac{1}{\sqrt{2}}
  \cph{\varphi_2}, \quad z= \frac{1}{\sqrt{2}} \cph{\varphi_3}, \quad  w =
   \frac{1}{\sqrt{2}} \cph{\varphi_4}
\end{aligned}
\]
where $\varphi_1 = 1.1051$, $\varphi_2 = -1.7482$, $\varphi_3 = 2.9750$,
$\varphi_4 = 0.9596$, $\theta_{\tau}$ are linearly spaced in the interval
$[0,2\pi]$, and $\tau = 1,2,\dots,40$. Notice how the variation among the
\num{40} surrogate models displayed in Fig.~\ref{fig:testFig}, for $N_s =
3,9,12$, decreases as the number of available samples increases.

When the method proposed here is used in the benchmarking cases of Section
\ref{sec:results}, $\num{6}$ randomly generated matrices of appropriate size
are used (\num{3} $\mathbf{G}_1$ matrices and \num{3} $\mathbf{G}_2$ matrices). 
This method has been tested on \num{6} and \num{90} randomly generated matrices,
with no observed significant change in performance observed across the tests.
Hence, for the practical implementation, \num{6} matrices are generated in
MATLAB, with \texttt{2*rand($p$,$m$)-1}, where $p$ and $m$ are the number of
rows and columns of the approximated frequency response.

\subsection{Loewner Generating System Based Adaptive Sampling II}
\label{sec:ThetaII}
The here proposed method uses the condition number of the block matrix function
$\mathbf{\Theta}(s)$ (or $\bar{\mathbf{\Theta}}(s)$) as a function of discrete
frequency to sample the next frequency point. 

The frequency-dependent condition number with respect to the $\ell^2$-norm is
calculated as,
\begin{equation}
  \kappa(\mathbf{\Theta},s_{\ell}') = ||\mathbf{\Theta}(s_{\ell}')||_2
  ||\mathbf{\Theta}(s_{\ell}')^{-1}||_2 = ||\mathbf{\Theta}(s_{\ell}')||_2
  ||\bar{\mathbf{\Theta}}(s_{\ell}')||_2 \text{.}
  \label{eq:conditionNum}
\end{equation}
With the condition number as a function of discrete frequency,
$\kappa(\mathbf{\Theta},s_{\ell}')$, we sample next frequency point according to
the lowest corresponding condition number, \ie, 
\begin{equation}
  s_{N_s+1} = \argmin_{s_{\ell}'} \kappa(\mathbf{\Theta},s_{\ell}') \text{.}
  \label{eq:conditionNumRule}
\end{equation}
To efficiently calculate $\mathbf{\Theta}(s_{\ell}')$ and
$\bar{\mathbf{\Theta}}(s_{\ell}')$, consider the identities.
\[
  (s_{\ell}'\mathbb{L} - \mathbb{L} \mathbf{\Lambda})^{-1} = (s_{\ell}'\mathbb{I} -
  \mathbf{\Lambda})^{-1}\mathbb{L}^{-1} \text{,}
\]
and
\[
  (s_{\ell}'\mathbb{L} - \mathbf{M}\mathbb{L}) ^{-1} = \mathbb{L}^{-1}
  (s_{\ell}'\mathbb{I} -
  \mathbf{M})^{-1}\text{,}
\]
where $\mathbf{\Lambda}$ and $\mathbf{M}$ are diagonal matrices, given in
\eqref{eq:LoewnerRightData} and \eqref{eq:LoewnerLeftData}, respectively, for
points such that $\mathbb{L}$ is invertible.

\section{Numerical Examples}
\label{sec:results}
In this section, we apply vector fitting and the Loewner framework to antenna
frequency responses using both predetermined and adaptive frequency sampling
algorithms. Here, the goal is to find the smallest number of frequency samples
necessary to reach a certain \gls{RMSE} tolerance.
%
%
%
%

When we use the Loewner framework, the surrogate models are on the form
\eqref{eq:LoewnerInterpolant1}, with state variables calculated from
\eqref{eq:interpolant2StateSpace}, with $\mathbf{D} = 0$.

Due to the Theta I algorithm's dependence on a set of \num{6} random matrices,
the method is tested \num{30} times (each time with a new set of \num{6} random
matrices, in \eqref{eq:LoewnerGenerating}) in every benchmark case. In each
benchmarking graph, the \gls{RMSE} span of the method is highlighted with the
yellow shaded area. The median of the \num{30} \gls{RMSE} curves is highlighted
with a solid yellow curve. \revision{Additionally, for the displayed
approximated relative errors using \eqref{eq:errorApprox}, a larger number of
random matrices are used (typically \num{300}). This large number of random
matrices is, however, only for displaying the estimated error, and not for
driving the choice of which frequency points to choose.}

In each antenna benchmarking case, equidistant distributions are
tested with the Loewner framework and vector fitting. These distributions are
equidistant for each iteration and are not iteratively enriched from previous
iterations.

\subsection{5G-antenna Example}
\label{sec:5G}
\begin{figure}[ht!]
  \begin{center}
    \includegraphics[width=1\columnwidth]{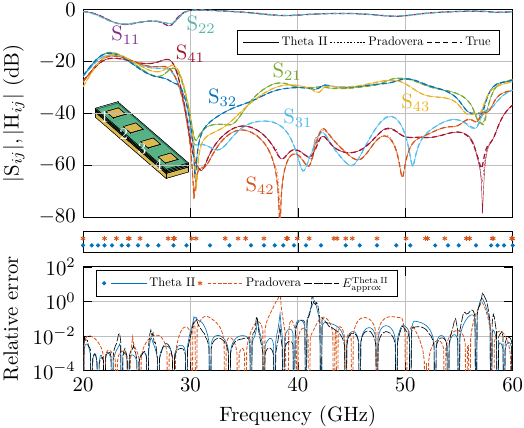}
  \end{center}
  \caption{In the top plot, the simulated frequency response ('true') of the
    5G-antenna is displayed, as well as the surrogate models constructed using
    $N_s = 34$ frequency points sampled according to the Theta II and Pradovera
    algorithm. The middle plot shows the corresponding frequency distribution.
    In the bottom plot, the relative errors \eqref{eq:error} of the surrogate
    models are displayed, including the approximated error
\eqref{eq:errorApprox}.}\label{fig:5G_antenna_ex}
\end{figure}

In this example, we consider the frequency response of a 5G-antenna, see inset
in Fig.~\ref{fig:5G_antenna_ex}, available in CST microwave studio's component
library \cite{CST}. The antenna response is calculated for \num{400} samples in
the range \SI{20}{\giga\hertz} to \SI{60}{\giga\hertz}, using CST's \gls{MoM}
solver, depicted in Fig.~\ref{fig:5G_antenna_ex} by the dotted line marked
'True'. \revision{In order to include a rapidly changing out-of-band data set to
  test the methods on, the frequency band of the simulation has been extended
beyond the bandwidth of the antenna.}

\begin{figure}[ht!]
  \begin{center}
    \includegraphics[width=1\columnwidth]{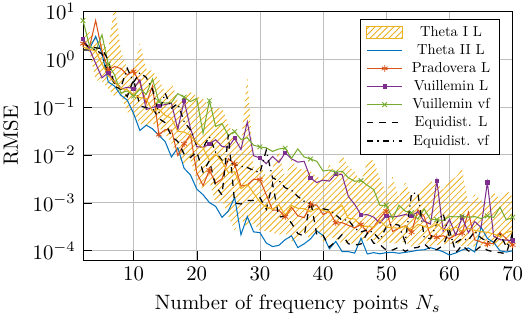}
  \end{center}
  \caption{\gls{RMSE} \eqref{eq:RMSE} obtained by letting the adaptive and
  predetermined frequency sampling algorithms use \num{2} to \num{70} frequency
  points, for the 5G antenna case.}\label{fig:5G_antenna_bench}
\end{figure}
\revision{
In Fig.~\ref{fig:5G_antenna_bench}, the \gls{RMSE} values obtained for the
adaptive sampling methods and the equidistant distributions are shown. We
observe in Fig.~\ref{fig:5G_antenna_bench} that the Loewner framework reaches a
consistently lower \gls{RMSE} compared to vector fitting. The lowest \gls{RMSE}
is achieved by the here proposed $\lTheta$ adaptive algorithm. The $\lTheta$
algorithm is the only algorithm in Fig.~\ref{fig:5G_antenna_bench}
to distinctly outperform the equidistant distribution for the Loewner framework.
Among the adaptive sampling algorithms, $\lTheta$ yields a consistently lower
\gls{RMSE}.}


The surrogate models generated using the Loewner framework yield the results
displayed in Fig.~\ref{fig:5G_antenna_ex}. \revision{In order to improve
readability of the plot, only a selection of scattering parameters have been
plotted}. For this case, \revision{\num{34}} frequency points have been
sampled using the Pradovera and $\lTheta$ algorithms, respectively. By sampling
according to $\lTheta$, the obtained overall relative error is lower, at most
frequencies, than compared to sampling with the Pradovera algorithm, and all
other tested adaptive methods. 
\revision{Additionally, the error of the Theta II generated surrogate model is
accurately approximated using \eqref{eq:errorApprox} with $N = 300$.} The
obtained sampling distributions using the two adaptive methods differ
noticeably. A denser distribution of samples is obtained where the coupling
terms are greater in magnitude for the $\lTheta$ algorithm.

\subsection{7$\times 1$-Vivaldi Array}
\label{sec:VivaldiArray}
\begin{figure}[ht!]
  \begin{center}
    \includegraphics[width=1\columnwidth]{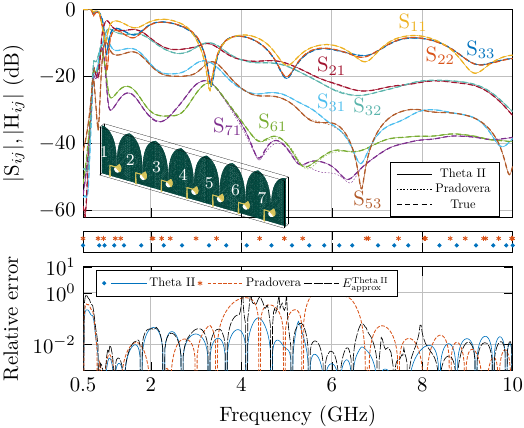}
  \end{center}
  \caption{In the top plot, the \gls{MoM} frequency response ('true') of the
    Vivaldi array is displayed, as well as the surrogate models constructed
    using $N_s=26$ frequency points sampled according to the Theta II and
    Pradovera algorithm. The middle plot shows the corresponding frequency
    distribution.  In the bottom plot, the relative errors of the surrogate
    models are displayed, including the approximated
    error.}\label{fig:vivaldiTestPlot}
\end{figure}

For our next example, the frequency response from a $7\times 1$ Vivaldi Array is
used, see inset in Fig.~\ref{fig:vivaldiTestPlot}. The frequency response is
calculated for \num{400} samples in the range \SI{0.5}{\giga\hertz} to
\SI{10}{\giga\hertz}, using an in-house \gls{MoM} solver, and is illustrated by
the dotted line in Fig.~\ref{fig:vivaldiTestPlot}.
\begin{figure}[ht!]
  \begin{center}
    \includegraphics[width=1\columnwidth]{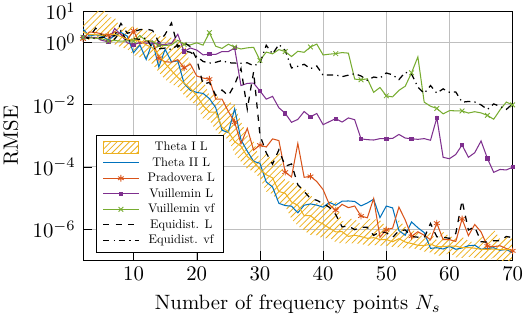}
  \end{center}
  \caption{\gls{RMSE} obtained by letting the adaptive and predetermined
    frequency sampling algorithms use \num{2} to \num{70}
  frequency points, for the $7\times 1$-Vivaldi array.}
  \label{fig:Vivaldi_bench}
\end{figure}

Figure~\ref{fig:Vivaldi_bench} displays the \gls{RMSE} from feeding the adaptive
and predetermined methods \num{2} to \num{70} frequency points. Observing the
yellow and blue lines, the corresponding methods, Theta I and II, yield similar
results in terms of \gls{RMSE}. The two methods perform similarly between
\num{2} to \num{35} frequency points. From \num{35} to \num{57} frequency
points, the Theta I method yields, on average, a lower \gls{RMSE}. Up until
\num{40} frequency points, both Theta I and II outperform the equidistant
distribution for the Loewner framework, whereas after \num{40} frequency points,
only the Theta I algorithm consistently outperforms the equidistant
distribution.


In Fig.~\ref{fig:vivaldiTestPlot}, the surrogate models generated using the Loewner
framework with \revision{\num{26}} frequency points sampled according to
the Pradovera algorithm and Theta II are shown. \revision{Only a small
selection of scattering parameters are plotted to improve visibility. Among the
selected scattering parameters, the most difficult to approximate are included}.
The corresponding relative errors between the true frequency response and the
surrogate models, \revision{as well as the approximated error}, are plotted below
the sample distribution of the two methods.

\revision{
The relative error corresponding to the Pradovera algorithm is observed to be
considerably fluctuate compared to the relative error corresponding to the Theta
II algorithm. 
Near \num{6} GHz for the $S_{71}$ parameter, the errors of the surrogate model
corresponding to the Pradovera algorithm become visible. The sensitive approach
of \eqref{eq:conditionNumRule} minimizes such occurrences for the here
proposed algorithm.
Apart from these errors, both the Pradovera and Theta II algorithm generated
models show good agreement with the true response. Additionally, the relative
error of the Theta II generated surrogate model is accurately approximated using
\eqref{eq:errorApprox} with $N = 300$.}


\subsection{8$\times$2 T-Slot Loaded Dipole Array}
\label{sec:LoadedDipoleArray}
\begin{figure}[ht!]
  \begin{center}
    \includegraphics[width=1\columnwidth]{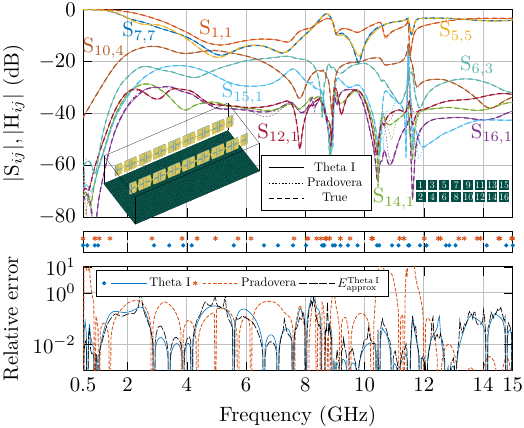}
  \end{center}
  \caption{In the top plot, the \gls{MoM} frequency response ('true') of the
    T-slot loaded dipole array is displayed, as well as the surrogate models
    constructed using $N_s = 35$ frequency points sampled according to the Theta I
    and Pradovera algorithm. The middle plot shows the corresponding frequency
    distribution.  In the bottom plot, the relative errors of the surrogate
    models are displayed, including the approximated
    error.}\label{fig:loadedDipoleTestPlot}
\end{figure}

For the next-to-last benchmarking example, we use the frequency response of a
$8\times2$ T-slot loaded dipole array, see inset of
Fig.~\ref{fig:loadedDipoleTestPlot}. The antenna element used in this example is
fully metallic, unlike the original T-slot loaded dipole element from
\cite{kolitsidas_rectangular_2014}. The frequency response is calculated using
an in-house \gls{MoM} solver for \num{400} samples in the range
\SI{0.5}{\giga\hertz} to \SI{15}{\giga\hertz}, and can be seen in Fig.
\ref{fig:loadedDipoleTestPlot}. The frequency band of the simulation has been
extended beyond the bandwidth of the antenna to provide rapidly changing
out-of-band data to test the methods on.

\begin{figure}[ht!]
  \begin{center}
    \includegraphics[width=1\columnwidth]{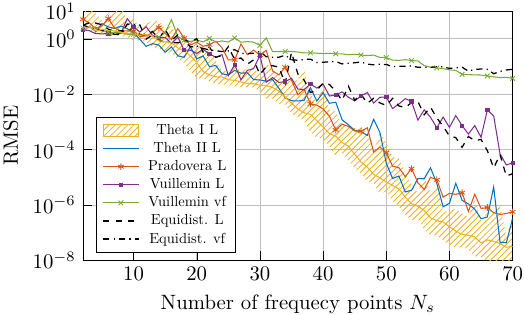}
  \end{center}
  \caption{\gls{RMSE} obtained by letting the adaptive and predetermined
    frequency sampling algorithms use \num{2} to \num{70}
  frequency points, for the $8\times 2$ T-slot loaded dipole array.}
  \label{fig:loadedDipole_bench}
\end{figure}

In Fig.~\ref{fig:loadedDipole_bench}, the \gls{RMSE} of the sampling methods is
displayed. The Theta I algorithm yields a noticeably smaller \gls{RMSE} for the
same number of frequency points used compared to the other methods. The
three adaptive methods, Theta I, Pradovera and Theta II, are observed to yield a
distinctly lower \gls{RMSE} than any predetermined distribution (after \num{40}
frequency points used). The difference in performance between the Loewner
framework and vector fitting is also clear.

Figure~\ref{fig:loadedDipoleTestPlot} shows the surrogate models generated from
using the Loewner framework with \revision{\num{35}} frequency points sampled
according to the Theta I and Pradovera algorithm. \revision{To increase
visibility, only a selection of scattering parameters have been plotted, among
them the most difficult to approximate.} In the lower plot of
Fig.~\ref{fig:loadedDipoleTestPlot}, the relative errors between the
surrogate models and the true frequency response are displayed, \revision{as
well as the approximated error}. \revision{Comparing the relative errors
corresponding to the two methods, the relative error corresponding to the
Pradovera algorithm is considerably greater in the \num{6.5} and \num{10.5} GHz
range, which yields visible errors in the model. These errors occur for
responses below \num{-40} dB in absolute magnitude. The relative error
corresponding to the Theta I algorithm is in contrast smoother. Additionally,
the relative error of the Theta I generated surrogate model is accurately
approximated using \eqref{eq:errorApprox} with $N = 300$.}


\subsection{4$\times$3 BoR Array}
\label{sec:BoRArray}
\begin{figure}[ht!]
  \begin{center}
    \includegraphics[width=1\columnwidth]{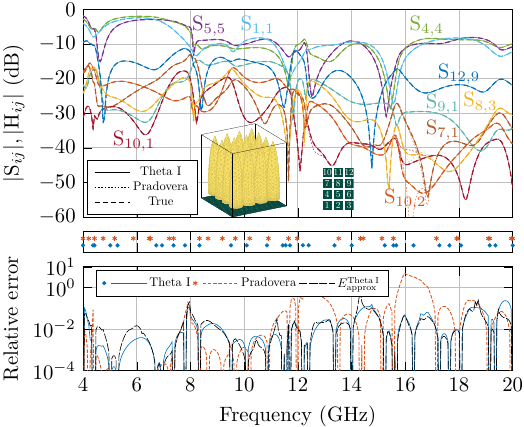}
  \end{center}
  \caption{In the top plot, the \gls{MoM} frequency response ('true') of the BoR
    array is displayed, as well as the surrogate models constructed using
    $N_s = 30$ frequency points sampled according to the Theta I and Pradovera
    algorithm. The middle plot shows the corresponding frequency distribution.
    In the bottom plot, the relative errors of the surrogate models are displayed,
    including the approximated error.}\label{fig:BoRArrayTestPlot}
\end{figure}

For the last benchmarking example, we use the frequency response of a $4\times3$
BoR array, see inset of Fig.~\ref{fig:BoRArrayTestPlot}, where the BoR element
is from \cite{holter_dual-polarized_2007}. The frequency response is calculated
for \num{300} samples in the range \SI{4}{\giga\hertz} to \SI{20}{\giga\hertz},
using an in-house \gls{MoM} solver. The frequency response of the BoR array is
depicted by the dashed lines in Fig.~\ref{fig:BoRArrayTestPlot}.
\begin{figure}[ht!]
  \begin{center}
    \includegraphics[width=1\columnwidth]{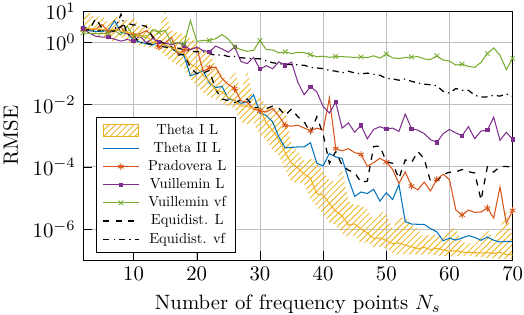}
  \end{center}
  \caption{\gls{RMSE} obtained by letting the adaptive and predetermined
    frequency sampling algorithms use \num{2} to \num{70}
  frequency points, for the $4\times 3$ BoR array.}
  \label{fig:BoRArray_bench}
\end{figure}

Figure~\ref{fig:BoRArray_bench} displays the \gls{RMSE} of the predetermined and
adaptive sampling methods for the BoR array case. Up until \num{30} frequency
points, the Theta I, Theta II and Pradovera algorithms perform similarly, or
better than the equidistant distribution for the Loewner framework. After
\num{30} frequency points used, the Theta I and the Theta II algorithms
distinctly outperform the equidistant distribution and the Pradovera algorithm.
Of the two, the Theta I algorithm yields the lowest \gls{RMSE} from \num{30} to
\num{70} frequency points used.


In Fig.~\ref{fig:BoRArrayTestPlot}, we display the surrogate models constructed
using the Loewner framework with \revision{\num{30}} frequency
points sampled according to the Theta I and Pradovera algorithm.
\revision{Only a selection of scattering parameters have been plotted in order
to increase visibility. Among the selected scattering parameters are the most
difficult to approximate.} The lower plot of Fig.~\ref{fig:BoRArrayTestPlot}
displays the corresponding relative error of the surrogate models
\revision{as well as the approximated error. The surrogate models corresponding
to the two methods show good agreement with the true data. Below \num{-40} dB
at \num{12.8} and \num{16} GHz, however, visible errors are observed from the
model corresponding to the Pradovera algorithm, whereas the model
corresponding to the Theta I algorithm shows no visible errors. Comparing the
relative errors of the surrogate models corresponding to the two methods, the
Theta I algorithm yields a relative error that is in contrast more smooth.
Additionally, the relative error of the Theta I generated model is
accurately approximated using \eqref{eq:errorApprox} with $N = 300$.}

\section{Conclusions}
\label{sec:conclusions}
We have proposed two novel adaptive frequency sampling algorithms, Theta I, and
II, based on the Loewner generating system approach. These adaptive frequency
sampling methods are used in conjunction with the Loewner framework and have
been compared with other frequency sampling algorithms in four antenna
responses. Among the adaptive methods, the Theta II, Theta I, and the Pradovera
algorithm succeed in sampling a minimal number of frequency points while still
achieving a remarkably low \gls{RMSE} in the resulting surrogate models,
demonstrating their performance in maintaining accuracy with reduced
computational effort. Additionally, in all of our benchmarking cases, the
Loewner framework has been shown to provide the most accurate surrogate models
compared to vector fitting, for all frequency sampling strategies tested in this
paper.

Of the three adaptive algorithms, Theta II yields the lowest average \gls{RMSE}
in one of the four antenna cases (the 5G antenna), whereas the Theta I algorithm
yields the lowest average \gls{RMSE} in three of the four antenna cases (The
Vivaldi array, the loaded dipole array, and the BoR array).


The proposed Theta I algorithm is appropriate for adaptive sampling for
minimizing an error of choice. In this paper, we have used an element-wise
relative difference calculation, leading to a minimization of the largest
element-wise relative error. The effect of this can be seen by the smoothness of
the relative error in the relative error plots of some of the benchmarking
cases. 


We have also shown that the proposed adaptive sampling algorithms perform better
than the tested predetermined frequency sampling distributions. With the
proposed adaptive sampling algorithms together with the Loewner framework, the
necessary number of frequency points can be kept low while still maintaining an
accurate system model.



\section*{Acknowledgment}
This work is supported by project nr ID20-0004 from the Swedish Foundation for
Strategic Research and \#2022- 00833 in the Strategic innovation program Smarter
Electronics System, by Vinnova, Formas, Energimyn- digheten
(Energy Agency), and the Swedish Research Council's Research Environment grant
(SEE-6GIA 2024- 06482) for research on sixth-generation wireless systems (6G),
which we gratefully acknowledge.

\ifCLASSOPTIONcaptionsoff
  \newpage
\fi


\bibliographystyle{IEEEtran}
\bibliography{./bibtex/bib/CompleteRef2,./bibtex/bib/CST_microwave}

\end{document}